\documentclass[preprint,12pt]{myclass}

\usepackage{fullpage}
\usepackage{graphicx}
\usepackage{multirow}
\usepackage{amsmath}
\usepackage{amssymb}
\usepackage[nodots,nocompress]{numcompress}
\usepackage[colorlinks,hyperindex]{hyperref}
\begin{document}

\begin{frontmatter}



\title{Effect of SiO$_2$ coating in bolometric Ge light detectors for rare event searches}

\author[BER]{J.W.~Beeman}
\author[ORS]{A.~Gentils}
\author[INS,ORS]{A.~Giuliani}
\author[INS]{M.~Mancuso}
\author[BIC]{G.~Pessina}  
\author[ORS]{O.~Plantevin}
\author[INS]{C.~Rusconi}

\address[BER]{Lawrence Berkeley National Laboratory, Berkeley, California 94720, USA}
\address[ORS]{Centre de Spectrom\'etrie Nucléaire et de Spectrom\'etrie de Masse, CNRS and Universit\'e Paris-Sud, F-91405 Orsay, France}
\address[INS]{Universit\`a dell'Insubria, Dipartimento di Scienza e Alta Tecnologia, 22100 Como, Italy}
\address[BIC]{Universit\`a di Milano-Bicocca, Dipartimento di Fisica, and INFN, Sezione di Milano Bicocca, 20126 Milano, Italy}

\begin{abstract}
In germanium-based light detectors for scintillating bolometers, a SiO$_2$ anti-reflective coating is often applied on the side of the germanium wafer exposed to light with the aim to improve its light collection efficiency. In this paper, we report about a measurement, performed in the temperature range $25-35$~mK, of the light-collection increase obtained thanks to this method, which resulted to be of the order of 20\%. The procedure followed has been carefully selected in order to minimize systematic effects. The employed light sources have the same spectral features (peaking at $\sim 630$~nm wavelength) that will characterise future neutrinoless double beta decay experiments on the isotope $^{82}$Se and based on ZnSe crystals, such as LUCIFER. The coupling between source and light detector reproduces the configuration used in scintillating bolometers.  The present measurement clarifies the role of SiO$_2$ coating and describes a method and a set-up that can be extended to the study of other types of coatings and luminescent materials.
\end{abstract}


\end{frontmatter}

% \linenumbers

\section{Introduction and motivations}
\label{sec:intro}
The searches for the neutrino mass and for dark matter in the Universe are at present two of the most relevant and exciting fields in cosmology and particle physics. Crucial experiments in these areas require the detection of very rare events, such as the neutrinoless double beta decay (DBD)~\cite{DBD-review}, a rare nuclear transformation, and the nuclear recoils induced by weakly interacting massive particles (WIMPs)~\cite{WIMP-review}, viable candidates to the composition of the galactic dark matter (DM). Although in these two searches the signals are expected in two very different energy regions - of the order of few MeV for the former, less than 100 keV for the latter - the experimental approaches to these investigations often share common technological issues. In both cases, bolometers, for their superior energy resolution and low-energy sensitivity, are suitable devices for many experiments, like CDMS~\cite{CDMS}, EDELWEISS~\cite{EDELWEISS}, CRESST~\cite{CRESST} and ROSEBUD~\cite{ROSEBUD} in the DM field, or Cuoricino~\cite{Cuoricino}, CUORE~\cite{CUORE}, LUCIFER~\cite{LUCIFER} and AMoRE~\cite{AMORE} as for DBD. Bolometer-based neutrinoless DBD and DM experiments require extremely low levels of background, especially that arising from radioactive contaminants in the bolometers themselves and surrounding materials. 

Luminescent bolometers are double-readout devices able to measure simultaneously the phonon and the light yields after a particle interaction in the detector. This operation allows in some cases to tag the type of the interacting quantum, crucial issue for background control~\cite{Mestres, MI-CaF2, pirro-scint}. Light is currently emitted through scintillation (in this case one speaks of {\em scintillating bolometers}) and detected by a thin auxiliary bolometer facing the main one. In non-scintillating materials relevant for DBD, the particle identification can be achieved through the detection of the much weaker Cherenkov light~\cite{Taba, Roma-Cer}. Scintillating bolometers are presently exploited in dark matter searches such as CRESST~\cite{CRESST} and ROSEBUD~\cite{ROSEBUD}, and in DBD searches such as LUCIFER~\cite{LUCIFER}, AMoRE~\cite{AMORE}, the future project EURECA~\cite{EURECA}, and an R\&D research line using ZnMO$_4$ as bolometric material~\cite{ZMO1, ZMO2}.

The crucial point is that a massive charged particle can be separated from an electron or a $\gamma$ due to the different light yield for the same amount of deposited heat. In DM searches, it is possible then to reject background due to ordinary $\beta$ and $\gamma$ radioactivity, which delivers energy to a primary electron, versus WIMP interactions, which takes place with nuclei. In neutrinoless DBD, the signal consists of a peak in the two-electron energy spectrum placed at the $Q$-value of the decay. For high $Q$-value candidate ($> \sim 2.5$~MeV) energy-degraded $\alpha$'s produced at the surface of the materials represent the most problematic background source. As for nuclear recoils, the scintillation yields of $\alpha$'s is generally lower than for $\beta$'s, with the significant exception of ZnSe. In all cases, an $\alpha$ event can be separated by a $\beta$ or $\gamma$ event (or in particular by a DBD event) looking at the ratio between scintillation and phonon yield.

The sensitivity required for the light detectors, in terms of total absorbed energy, depends on the type of search, on the material used for the main bolometer and on the light-collection efficiency. In the case of DM, where very low-energy events are searched for, the scintillation photons produced by a background event are always very few, and the corresponding scintillation energy is of a few tens of eV only: this fixes the performance requested in terms of energy threshold. In the case of DBD, one spans from the less problematic case of the scintillator ZnSe (used for the study of the isotope $^{82}$Se having a $Q-$value of 2996~keV), which has shown a sample-dependent light yield of the order of 10-30 keV/MeV for $\alpha$'s and approximately 3-4 times less for $\beta$'s~\cite{Mi-ZnSe}, to the very challenging situation of Cherenkov light emission, to be exploited for non-scintillating material such as TeO$_2$ (used for the study of the isotope $^{130}$Te having a $Q-$value of 2527~keV), that corresponds to only 60~eV/MeV~\cite{Roma-Cer}. The scintillator ZnMO$_4$ (used for the study of the isotope $^{100}$Mo having a $Q-$value of 3034~keV) represents an intermediate case, with a light yield of the order of 1 keV/MeV for $\beta$'s~\cite{ZMO1, ZMO2}. Therefore, the hypothetical peak of neutrinoless DBD would appear in the light detector at about 15-25 keV for ZnSe, at about 3 keV for ZnMoO$_4$ and only at 140 eV for TeO$_2$. In the last case, the performance required for DBD and DM searches are similarly challenging.   

In this work, we refer to bolometric light detectors that use high-impedance neutron transmutation doped (NTD) Ge thermistors as temperaure sensors~\cite{NTD-haller}. This method for the read-out of the thermal signal is quite popular (it is used in the DBD searches CUORE~\cite{CUORE} and LUCIFER~\cite{LUCIFER} and in the DM searches EDELWEISS~\cite{EDELWEISS} and ROSEBUD~\cite{ROSEBUD}), due to the simplicity of the detector assembly and operation and to the use of a conventional front-end electronics. The light detectors consist of hyper-pure Ge thin slabs equipped with NTD thermistors. This structure is convenient and very well tested~\cite{ROSEBUD, pirro-scint}. However, due to the the very low energy of the expected signals, an improvement of the light-detector sensitivity, which is crucial to achieve an adequate background rejection, is mandatory for the most challenging situations (DM application and TeO$_2$) and in any case welcome for the ZnSe and ZnMoO$_4$ cases. The sensitivity improvement of the Ge light detectors here considered can be obtained following four lines: (i) optimization of the NTD thermistor parameters; (ii) reduction of the heat capacity of the energy absorber making it thinner (present thickenesses range from 0.1 to 1 mm); (iii) decrease of the operation temperature from the present 15-20 mK to 10 mK; (iv) use of proper anti-reflective coatings of the Ge side exposed to the luminescent bolometer. The present article deals with the last point, proving and quantifying the positive effect of a SiO$_2$ coating.  Since the work here described was performed in the framework of the LUCIFER project, the scintillating material employed in the tests is ZnSe. However, the presented method and set-up can be easily adapted to investigate other luminescent crystals and different coating materials.

\section{Description of the method and of the set-up} 
\label{sec:setup}
The performance of NTD-based bolometers is on the average quite satisfactory and has allowed and allows to run very competitive experiments, such as MiBeta~\cite{MiBeta}, Cuoricino~\cite{Cuoricino} and EDELWEISS~\cite{EDELWEISS}. However, the reproducibility of these devices remains a problem to be solved. It is not unusual that nominally identical detectors differ by a factor two or more in energy-voltage conversion, energy resolution and threshold. These differences are due to the difficult reproducibility of thermal couplings at very low temperatures and to the detector dependence of spurious noise sources, like microphony and erratical heating induced by vibrations. Therefore, it is never obvious to pick up the effect of a single constructive parameter on the bolometer performance. We had this in mind when we designed a set-up to investigate the effect of SiO$_2$ coating.

The main idea was to fabricate thin detectors with a coating on only one side, and to make them as symmetric as possible with respect to the two sides in any other aspect. The same light source was then used to illuminate each light detector, in a first cryogenic run from one side and in a second cryogenic run from the other side. The detector-source geometrical coupling was identical in the two runs. The bolometers were equipped with a resistive heater to normalize the detector performance in the two runs. (The same operation can be done, but with lower statistics, studying the detector response to single ionizing particles.) This normalization is mandatory if one wants to be sensitive to possibly tiny effects, since response differences from run to run are normally observed in NTD-based bolometers. In order to get a redundant confirmation of the results and to overcome inevitable small side asymmetries in the detector configuration, four bolometers have been realized.

\subsection{The light detectors}
\label{sec:ld}
Each detector has an energy/light absorber consisting of a hyperpure Ge square plate, with a side of 15 mm and a thickness of 0.3 mm. The four plates have been obtained from the same large Ge single crystal, provided by Ortec and characterized by an impurity concentration less than $10^{11}$cm$^{-3}$ and by a room temperature resistivity of $40$~$\Omega \cdot$~cm. After polishing and etching, a coating was made on one side of the plate by RF sputtering a SiO$_2$ target using argon with 10\% oxygen as a sputter environment.  The deposition was carried on for 2 hours at at 200~W forward power, which should produce a layer approximately 70~nm thick. No room temperature characterization of the film thickness and structure was made.

\begin{figure}[h!]
\centering
\includegraphics[width=0.5\textwidth]{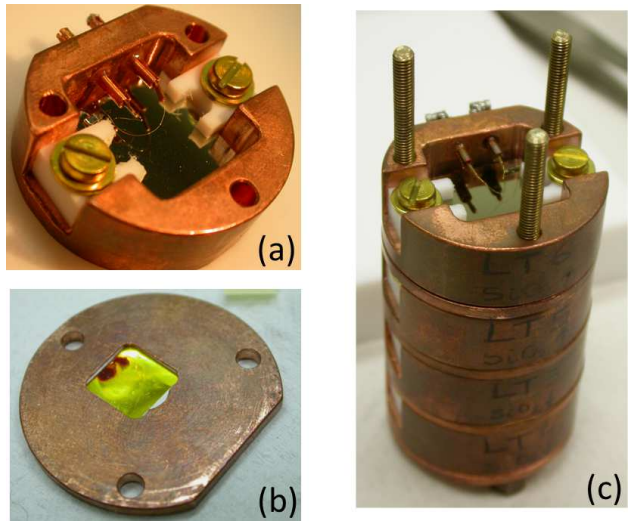}
\caption{(Color online) (a) Photograph of an individual light detector; (b) Photograph of the light source; (c) Photograph of the light-detector tower.}
\label{fig:det-photo}
\end{figure}

The Ge plate was inserted in a copper holder with approximateley the shape of a flat ring, with a thickness of 11 mm, an external diameter of 32 mm and a central square window (side 17~mm) to accomodate the plate, which is held by two opposite PTFE clamps establishing a firm mechanical coupling and a week thermal link towards the copper support (which acts as a heat sink for the device). The plate plane is centered with respect to the copper ring thickness. A photograph of an individual detector is shown in Fig.~\ref{fig:det-photo}(a).

The thermal signals from the Ge plate were provided by 3~$\times$~1~$\times$~0.6~mm NTD Ge thermistors, showing a resistivity-temperature law well compatible with the Variable Range Hopping conduction regime with Coulomb gap, which foresees:
\begin{equation}
\label{eq:VRH}
R (T)= R_0~exp[(T_0/T)^{0.5}]
\end{equation}
where $T_0$=3.83~K and $R_0$=6.45~$\Omega$ for the samples here used.  The thermistors were thermally connected to the energy absorber through the two-component epoxy {\sl Araldite rapide}. In order to well define the contact area with the energy absorbers, the thermistors presented a 50~$\mu m$ thick stand-off, with $1 \times 1$~mm area, placed at the center of one of the $3 \times 1$ face. The epoxy was placed in the form of a small dot on this stand-off, and then pressed in order to form a thin veil not overflowing from the protruding $1 \times 1$~mm region. 
A heater, used to stabilize the detector response and to fix a standard excitation allowing to compare results from different runs, was glued on the opposite side of the Ge plate with a single epoxy dot. As described elsewhere~\cite{heater}, it consists of a meander of heavily doped silicon obtained by ion implantation on a $3 \times 3$~mm wide and 0.5~mm thick Si chip. 
The thermistor and the heater were electrically connected by means of golden wires with a diameter of 50~$\mu m$ and about 15~mm long. The wires were ball-bonded at the thermistor and heater pads at one end and crimped in a copper tube at the other end. The tube was thermally anchored at the copper ring-shaped holder but electrically isolated from it, and was connected to high resistivity cryogenic wires for the signal read-out. 

Given the different shapes of the heater and of the thermistor and given the different effect that they would have on the thermal signal when illuminated by a light source (the thermistor would respond to it providing a fast additional component to the overall thermal signal, while the heater would shield the light providing on the contrary a slow component), the two elements were glued at a corner of the plate, in order to minimize the fraction of light reaching them.  

The thermistors and the heaters used for the four detectors are nominally identical. In two detectors, the thermistors were glued on the coated side, while in the other two ones on the bare side. This is the only intentional difference between the four detectors. It was introduced to exclude or at least control a possible effect determined by the inevitable asymmetry due to the presence of different elements on the two sides. From now on, the detectors with the thermistors on the coated side will be referred to as LD\#1 and LD\#2, while the remaining two as LD\#3 and LD\#4.

% LD2 -> LT5 - 2nd from above - S 1.3 -> coating in Run1 - the warmest
% LD1 -> LT6 - 1st from above - S 8.1 -> coating in Run1 - the second coldest
% LD3 -> LT3 - 3rd from above - S 4.3 -> coating in Run2 - the best and the coldest
% LD4 -> LT4 - 4th from above - S 8.2 -> coating in Run2 - the second warmest (varying)

\subsection{The light sources and the source-detector coupling}
\label{sec:sou}

The light sources (one of them is depicted in Fig.~\ref{fig:det-photo}(b)) were realized by sending ionizing radiation on four ZnSe slabs obtained by crystalline samples provided by the Alkor Technology company of Saint Petersburg (Russia), and described elsewhere~\cite{dafinei}. On one side of each slab, one or two drops of an uranium standard solution were deposited and then dried by several-hour exposition to air. The radioactive source carrier is a water solution of 1\% wt of HNO$_3$ containing originally 973~$\mu$g/ml of uranium, but it was diluted to reduce the rate at a level acceptable for a bolometer. Independent measurements showed that the secular equilibrium of $^{238}$U was broken, and that the uranium appeared to have been taken from a depleted stock, since less than expected decays from $^{235}$U and $^{234}$U were observed. A source spectral characterization showed $^{238}U$ $\alpha$ particles at 4.15 MeV (B.R. 21\%) and 4.20 MeV (B.R. 79\%). A weak doublet from $^{234}U$ was also observed at 4.77 MeV (B.R. 71.4\%) and 4.72 MeV (B.R. 28.4\%), with an intensity about 10 times lower than that of the lower energy doublet. Of course, the $\alpha$ particles are fully absorbed by the ZnSe element, and only scintillation light is expected to come out of the opposite side of the slab. We do not expect that a source energy spectrum obtained through light pulses preserves the well defined energy structure of the source with two sharp peaks, since the reaction of the acid with ZnSe produces an opaque region, as appreciable in Fig.~\ref{fig:det-photo}(b), with consequent partial light absorption. In the actual energy spectrum, we expect therefore a smeared structure with a low-energy tail in which however the two main characteristic $\alpha$~energies of the source ($\sim 4.2$ and $\sim 4.7$ MeV) should be recognizable. The side where the source was deposited was covered with a reflective foil in order to increase the light emission.

The use of ZnSe as a source material ensures that the spectral features of the emitted light is similar to that expected in the LUCIFER experiment, which envisages the use of this compound as neutrinoless DBD source. We have measured in detail the temperature-dependent photoluminescence emission of a ZnSe crystal from which two of the four employed sources were obtained (those facing detectors LD\#1 and LD\#4), from room temperature to 12~K. The photoluminescense was stimulated by an Ar laser emitting at $488$~nm ($2.54$~eV). This excitation was chosen since the laser wavelength is close to the value ($483$~nm) that provides the maximum emission when studying the wavelength-dependence of the sample photoluminescence with a monochromatized Xe lamp. The photon energy is below the ZnSe energy gap (that is $\sim 2.7$~eV), but it is sufficient to excite the main luminescence mechanism (donor-acceptor pair recombination), which is responsible for light emission also in the case of UV excitation and scintillation induced by ionizing particles. 

\begin{figure}[h!]
\centering
\includegraphics[width=0.6\textwidth]{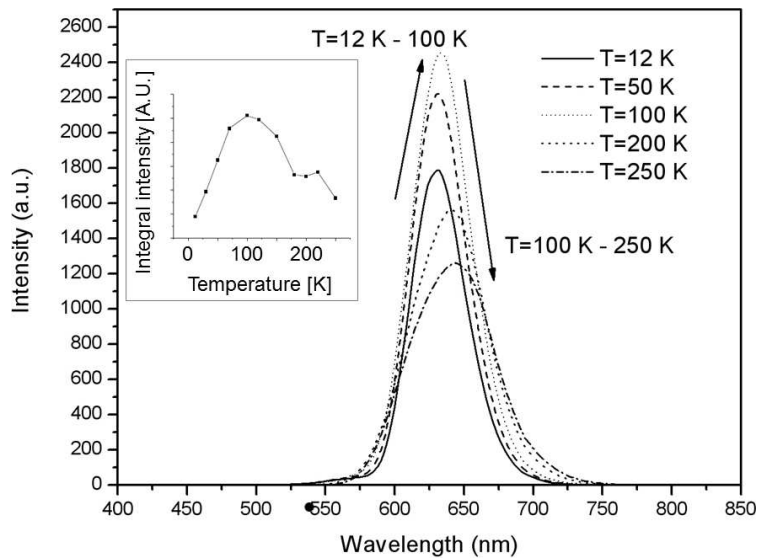}
\caption{Typical emission spectra collected at different temperatures of the light sources used to test the coating effect. The sources consist of ZnSe scintillating slabs excited by an Ar laser emitting at 488~nm. The shift of the peak position to higher wavelength as the temperature decreases is appreciable. In the inset, the integral intensity as a function of the temperature.}
\label{fig:emission}
\end{figure}

As shown in Fig.~\ref{fig:emission}, the emission spectrum is a broad peak extending from $\sim 550$ to $\sim 750$~nm. The maximum-emission wavelength varies smoothly from 641 nm to 632 nm when the temperature decreases from room temperature to 12~K. The light intensity initially increases by decreasing the temperature, reaching a maximum at about 100 K, where it is higher by a factor $\sim 1.5$ than at room temperature. It then decreases again down to a value at 12~K slightly lower than that at room temperature (see inset in Fig.~\ref{fig:emission}). 

Each ZnSe slab was inserted in a copper plate that covers the ring-shaped holder of the Ge detector (as visible in Fig.~\ref{fig:det-photo}(b)), keeping the light-emitting side of the source in front of the Ge plate, at a distance of about 5.5~mm. The ZnSe slab is slightly off-centered with respect to the Ge plate, in order to direct most of the light on a free Ge area, not covered by the heater, the thermistor and the related wires and copper pins for the contacts. The ensemble of these details can be appreciated in Fig.~\ref{fig:det-photo}. 

The four detectors and the four light sources were alternatively stacked in a tower stucture -- depicted in Fig.~\ref{fig:det-photo}(c) -- following the detector numbering, with LD\#1 above and closer to the mixing chamber. After a first cryogenic run in a given configuration (Run~\#1 from now on), a second run (Run~\#2) was performed after simply reversing each detector holder and changing therefore the detector side exposed to the source. A configuration was chosen so that in a given run the ligth sources illuminated the coated detector side in two cases and the uncoated ones in the other two. This choice allows to exclude or at least control possible pulse amplitude changes from Run~\#1 to Run~\#2 due to a modification of the cryogenic conditions, which could confuse the interpretation of the coating effect.

In both runs, a low-rate $^{55}$Fe source was placed in front of each light detector, illuminating the side opposite to that exposed to the light source. The $^{55}$Fe sources were obtained with a technique similar to that described for the uranium source, using an acid solution containing the radioactive isotope. The purpose of the $^{55}$Fe source, emitting two X-rays at 5.9 and 6.4~keV, is to provide an absolute energy calibration of the light detectors.

\subsection{Preliminary consideration on the coating effect}
\label{sec:coat}

The reflectance $R$ at the surface between a transparent medium and an absorbing one for light incident perpendicularly is calculable in the framework of geometrical optics and is given by the formula
\begin{equation}
\label{eq:reflect}
R = \frac{(n_0-n_1)^2+k_1^2}{(n_0+n_1)^2+k_1^2}
\end{equation}
where $n_0$ is the real refraction index of the transparent medium and $n_1$ and $k_1$ are respectively the real and the imaginary part of the complex refraction index of the absorbing medium. For germanium, which is absorbing in the optical range, n$_1$=5.48 and k$_1$=0.823 at 632~nm and at room temperature~\cite{dati-Ge}. These values have a weak wavelength dependence (that will be neglected in this discussion) in our relevant spectral range, corresponding to the emission spectrum of ZnSe. In case of vacuum-germanium interface the absorbed light intensity (given by $1-R$) is of the order of only 51\%. Of course, in a luminescent bolometer a part of the reflected light will reach the light detector again after reflections in the cavity containing the main crystal. However, this mechanism is not very efficient and an increase of the absorbed fraction would be very useful if a tiny luminescence energy has to be detected.

The simplest form of antireflective coating is based on the so-called index matching. If a thin layer of material with an intermediate refraction index $n_i$ is interimposed between the vacuum and the absorbing medium, then we have to apply twice the equation~(\ref{eq:reflect}), first from vacuum to the new medium (that will be chosen as transparent and therefore with $k=0$ in the wavelength range of interest) and then from the new medium to the absorbing one. There is an optimum value for $n_i$, that in case of germanium is 2.4. A layer with this feature would increase the absorbed fraction up to 69~\%, corresponding to a gain of 35~\% with respect to bare germanium. The material that we have deposited is SiO$_2$, which has a refractive index of 1.54 in crystalline form (quartz)~\cite{dati-Q} and 1.45~\cite{dati-FS} in amorphous form (fused silica), at room temperature and at 632~nm. These two values lead to an absorbed fraction of 64.5\% and 63.1\% respectively, which corresponds to a gain of 25.6\% and 22.7\%. The effect of our coating is however difficult to predict, because of uncertainties in the structure and stoichiometry of SiO$_2$ (which affects the refraction index) and in the temperature dependance of the refraction indices. In addition, the source is not collimated and quite close to the detector (as happens in scintillating bolometers): therefore, the assumption of normal incidence is a crude approximation. Multiple detector-source reflections  are not taken into account. The situation is complicated also by the fact that the light wavelength is an order of magnitude longer than the SiO$_2$ film thickness, while equation~(\ref{eq:reflect}) applies in principle to bulk materials. However, we can take the predicted gain in case of amorphous SiO$_2$ (22.7\%) as an indication of the improvement that can be obtained with the present method. As we will see in the Section~\ref{sec:results}, experimental results will not be far from this value. These results are significant, as the detector-source geometrical configuration reproduces faithfully the structure normally adopted in scintillating bolometers, presenting a gap of a few millimeters between the light-emitting element and the light detector, which have similar surface areas.

\section{Results and analysis}
\label{sec:results}

\subsection{Cryogenics, read-out and data acquisition}
\label{sec:cry-ele}
The detector tower described in Section~\ref{sec:sou} was installed both in Run~\#1 and in Run~\#2 in the experimental vacuum of a low-power dilution refrigerator located in the cryodetector laboratory of the University of Insubria, Como, Italy. The tower was thermally connected to the coldest point of the refrigerator, namely the mixing chamber. The nominal base temperature of the dilution refrigerator with no thermal load is 13~mK. The cooling power is around 25~$\mu$W at 100~mK.  

In order to read out the thermal pulses produced by light flashes and single particle interactions, each NTD Ge thermistor was supplied with an almost constant current, obtained by applying a total DC voltage bias $V_{tot}$ at the series of thermistor and two $100$~M$\Omega$ load resistances, placed at room temperature and connected symmetrically at the two thermistor sides in order to respect the differential configuration of the front-end electronics. The voltage signals across the thermistors were sent to room-temperature differential low-noise voltage amplifiers and to Bessel analogical filters with $125$~Hz cut-off frequency, operating as an antialiasing stage. The data acquisition was performed by means of a dedicated software, also controlling the trigger: waveforms were sampled at 20 kHz (1024 sample points in a 50 ms time window), digitized by a 12 bit LeCroy module and registered by a 500 MHz Intel processor. The ADC configuration allowed to trigger only one channel at once (acting as a common trigger on all the channels), so at least four separate measurements were performed for each run. The pulses of the non-triggered channels were anyway registered and used for coincidence analysis, very useful for the energy calibration described in Subsection~\ref{sec:ec}.

The heaters described in Section~\ref{sec:ld} were connected in parallel to a single wire pair from the mixing chamber to room temperature. A pulse generator was programmed in order to inject every few minutes a square voltage pulse $V$ with a duration of $1$~ms, shorter than the thermal signal risetime, into each of the four heaters (whose resistance is $\sim 300$ k$\Omega$ at the detector operation temperature) . The energy dissipated in the heater is trivially given by the Ohm's law and the Joule's power formula. The value calculated in this way cannot however be taken as an absolute detector calibration. The heaters were coupled to the heat sink with gold wires which introduce a non-negligible heat conductance with respect to that of the glue connecting the heating element to the germanium plate. Only a fraction of the energy is therefore dissipated in the light detector. The amplitude of the voltage pulse was tuned in order to provide a thermal pulse with an energy corresponding to $\sim 700$ keV on a calibrated energy scale. This value is higher enough to be out of all the detector spectral structures and low enough not to produce seriously deformed pulses due to strong non-linearity. The heater pulse represents an excellent method to compare energy spectra of Run~\#1 with those of Run~\#2 (see Subsection~\ref{sec:coating-effect}) and to fix a common energy scale. 

\subsection{Current-voltage curves and operation points}
\label{sec:iv}
\begin{figure}[h!]
\centering
\includegraphics[width=0.6\textwidth]{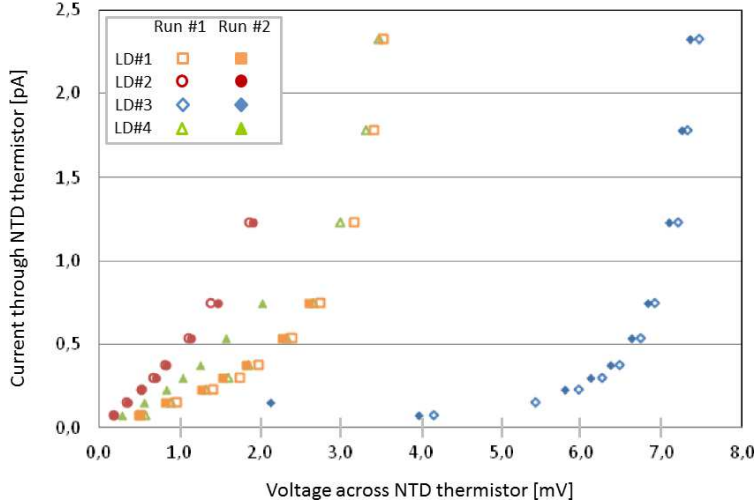}
\caption{(Color online) Current-voltage curves of the four light detectors in Run\#1 and Run\#2.}
\label{fig:iv}
\end{figure}

Prior to the measurements about the efficiency of the coating procedure, the static behaviour of each thermistor was studied. By varying the total DC bias $V_{tot}$ and measuring the corresponding voltages $V$ appearing across the thermistors through the low-noise amplifiers, a current-voltage ($I-V$) curve was collected for each detector in both runs. The eight $I-V$ curves (four in Run\#1 and four in Run\#2) are plotted in Fig.~\ref{fig:iv}. For three detectors, the $I-V$ curves of the two runs are quite similar. This means that the detector base temperature, the thermal couplings and the parasitic power due to vibrations and/or electrical effects have not substantially changed from Run\#1 to Run\#2 (although there is some indications that in Run\#2 the temperature was slightly higher, as pointed out later). On the contrary, the detector LD\#4 shows a remarkable higher temperature in Run\#2 and consistently a lower thermal coupling to the heat sink. This shows that the adopted detector structure can unfortunately lead to changes in the detector behaviour after thermal cycling.

Another apparent feature emerging from Fig.~\ref{fig:iv} is the diversity of $I-V$ curves (and of their temperatures at low biasas) for nominally identical detectors coupled to an isothermal heat sink. This confirms the irreproducibility problems mentioned at the beginning of Section~\ref{sec:setup}. It is to notice that there is no correlation between the temperatures of the detectors and their position in the tower. Therefore, the temperature dishomogeneity cannot attributed to a temperature gradient along the tower.  

The operation bias was chosen so as to maximize approximately the voltage signal amplitude. The parameters of the detectors at the chosen working points are summarized in Table~\ref{tab:wp}. In Run\#2, the common base temperature resulted a bit higher than in Run\#1, as it can be appreciated by the generally lower resistances for the same total DC bias.

\begin{table}[h]
\caption{Working points of the four light detectors in Run\#1 and Run\#2: $V_{tot}$ is the total DC bias, $V$ is the voltage across the NTD Ge thermistor, $R$ and $T$ its resitance and temperature respectively. The load resistance is always 200~M$\Omega$. The detectors are listed with increasing distances from the mixing chamber.}
\centering
\begin{tabular}{|c|rrrr|rrrr|}
\hline
%Detector & \multicolumn{3}{|c|}{Run\#1} & \multicolumn{3}{|c|}{Run\#2} \\
\multirow{2}{*}{Detector} & \multicolumn{4}{|c|}{Run\#1} & \multicolumn{4}{|c|}{Run\#2} \\
\cline{2-9}
& $V_{tot}$[V] & $V$[mV] & $R$[k$\Omega$] & $T$[mK] & $V_{tot}$[V] & $V$[mV] & $R$[k$\Omega$] & T[mK] \\
\hline
LD\#1 & 1.10 & 2.43 & 443 & 30.9 & 1.10 & 2.25 & 410 & 31.3 \\
LD\#2 & 1.10 & 1.14 & 208 & 35.5 & 1.10 & 1.08 & 197 & 35.9 \\
LD\#3 & 0.76 & 6.42 & 1704 & 24.6 & 0.30 & 5.00 & 3390 & 22.1 \\
LD\#4 & 1.10 & 2.37 & 432 & 31.0 & 1.10 & 1.65 & 301 & 33.1 \\
\hline
\end{tabular}
\label{tab:wp}
\end{table}

\subsection{Evaluation of the coating effect on light collection}
\label{sec:coating-effect}

For each detector and in both runs, data acquisition was performed at the reported working points without calibration sources placed outside the cryostat. The full waveforms of the signals from the NTD thermistors were acquired and registered, and the method of the optimum filter~\cite{OF} was applied in order to maximize the signal-to-noise ratio in an off-line analysis. 

\begin{figure}[h!]
\centering
\includegraphics[width=0.6\textwidth]{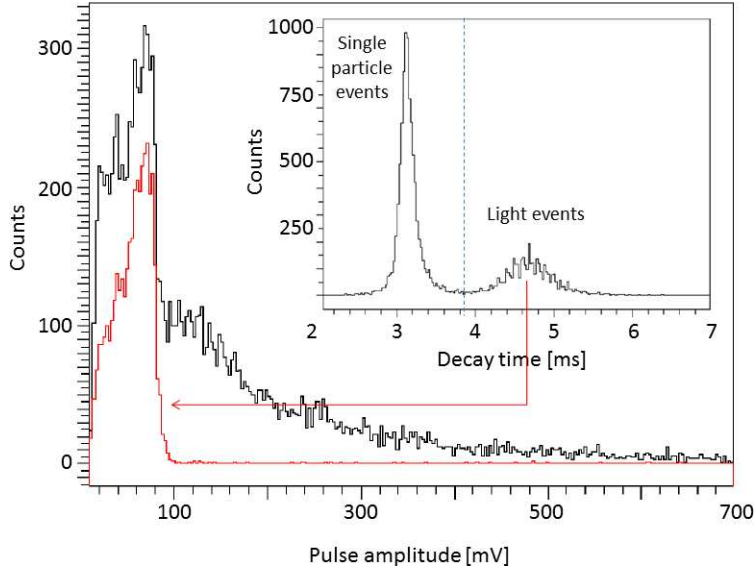}
\caption{(Color online) The total energy spectrum from detector LD\#3 is reported in black. The spectrum of the light signals (grey/red) is extracted thanks to pulse shape discrimination, using the decay time distribution (see inset). Similar results are obtained for the other light detectors.}
\label{fig:discr}
\end{figure}

Environmental radioactivity and cosmic rays provided a significant interaction rate in the light detectors, comparable to that due to the light sources, which resulted to be in the range $0.06 - 0.1$~Hz. As for the light sources, the rate was in agreement with the amount of uranium solution deposited on the ZnSe slabs. In order to isolate the component due to light flashes in the energy spectrum of the detectors, we have exploited the well known fact that the temporal structure of the signal is different for pulses due to light absorptions with respect to pulses due to single ionizing particles~\cite{Mi-ZnSe}. The separation is very efficient, as appreciable in Fig.~\ref{fig:discr}, and the light spectrum can be easily extracted from the total. Average rise times (10\%-90\% of pulse maximum) and decay times (90\%-30\% of pulse maximum) for both types of signals are reported in Table~\ref{tab:pulse}.

\begin{table}[h]
\caption{Pulse characteristics for the four light detectors. Rise time ($T_R$) and decay time ($T_D$) ranges are 10\%-90\% and 90\%-30\% of pulse maximum respectively. The detector sensitivity is based on the calibration performed with the $^{55}$Fe source and the cosmic rays (see Section~\ref{sec:ec}). Data refers to Run~\#1 and are similar to those obtained in Run~\#2.}
\centering
\begin{tabular}{|c|c|c|c|c|c|}
\hline
%Detector & \multicolumn{3}{|c|}{Run\#1} & \multicolumn{3}{|c|}{Run\#2} \\
\multirow{2}{*}{Detector} & \multicolumn{2}{|c|}{Single particle} & \multicolumn{2}{|c|}{Light} & Sensitivity \\
\cline{2-5}
& $T_R$ [ms] & $T_D$ [ms]  & $T_R$ [ms] & $T_D$ [ms] & [nV/keV] \\
\hline
LD\#1 & 2.06 & 3.64 & 2.29 & 5.24 & 470 \\
LD\#2 & 2.06 & 3.33 & 2.28 & 4.79 & 307 \\
LD\#3 & 2.27 & 3.18 & 2.47 & 4.65 & 523 \\
LD\#4 & 1.96 & 3.16 & 2.17 & 4.63 & 322 \\
\hline
\end{tabular}
\label{tab:pulse}
\end{table}

In order to compare the detector responses in the two runs (and therefore to investigate the effect of the coating), we have used the heater signals. We have adjusted the energy scales (in arbitrary units) so that the peaks due to the heater pulses are superimposed. This operation was not possible for the light detector LD\#1, since in Run~\#2 the contact to the heater opened. For this detector, the energy normalization was done using the shape of the energy spectrum of events due to single particle direct interactions, given mainly by cosmic muons crossing the detector (see Section~\ref{sec:ec}). 

\begin{figure}[h!]
\centering
\includegraphics[width=0.6\textwidth]{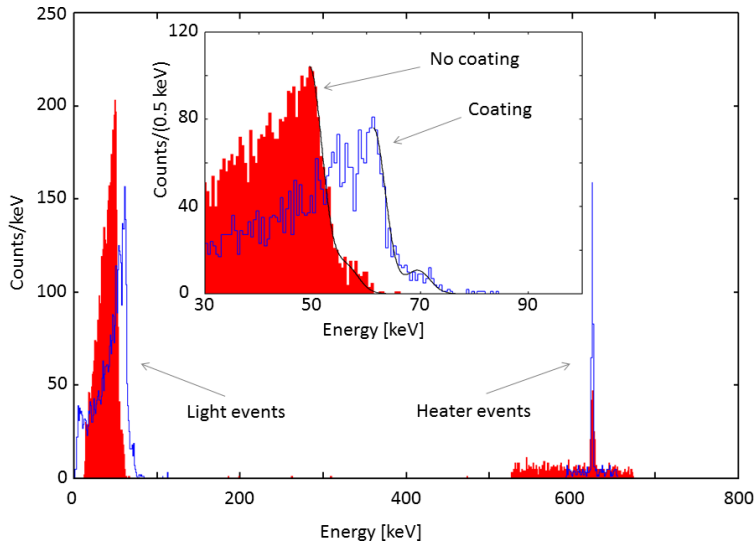}
\caption{(Color online) The energy spectra after selection of light pulses through pulse shape discrimination are reported for the detector LD\#3, both for Run~\#1 (red, filled histogram) and for Run~\#2 (blue, empty histogram), after normalization of the energy scale through the heater peak. The high-energy sides of the light spectra structures (corresponing to light generated by the alpha sources) are fitted with semi gaussians (see inset). The effect of the coating is apparent. Similar results are obtained for the other light detectors.}
\label{fig:comp}
\end{figure}

Two light spectra collected with the detector LD\#3 in the two runs are reported in Fig.~\ref{fig:comp} as an example. The gain due to the SiO$_2$ coating in terms of total energy asborbed by the light detector is apparent. Similar plots are obtained for the other detectors. It is confirmed that the SiO$_2$ coating gives reproducible positive effects on the light collection efficiency, which prevail over any other possible systematic factors that could simulate or fade this result.

The quantification of the achieved improvement was estimated by fitting the high-energy sides of the light spectra (not influenced by light absorption in the alpha source) with a sum of two semi-gaussian functions. The distance between the maxima of the two functions was fixed according to the positions of the two main lines of the uranium source described in Section~\ref{sec:sou}. The amplitude of the gaussian at higher energy resulted always about 10 times lower than the other gaussian, confirming a known feature of the employed alpha source. The position of the main maximum of the fitting function in Run~\#1 was compared with that in Run~\#2, taking into account their statistical uncertainties after energy normalization (which also contributes to the global uncertainty). The results are reported in Table~\ref{tab:results}, in terms of percentage improvement due to the SiO$_2$ coating. A large uncertainty characterizes the measurement of the improvement factor of detector LD\#1 due to the aforementioned failure of the heater contacts in Run~\#2, which made energy normalization much less precise. 

The dispersion of the improvement-factor values is consistent with the uncertainties of each measurement. This shows that the set-up has introduced no systematic effect that can affect the evaluation of the improvement factor, at least at the level of our sensitivity. The average improvement factor results to be $(20.1 \pm 0.6)$~\%, not far from the approximate estimation done in Section~\ref{sec:coat}.

\begin{table}[h]
\caption{Coating-induced improvement factors in the four light detectors analyzed in this work.}
\centering
\begin{tabular}{|l|c|c|c|}
\hline
Detector & Improvement factor [\%] \\
\hline
LD\#1 & $18.5 \pm 4.0$ \\
LD\#2 & $19.8 \pm 2.2$ \\
LD\#3 & $20.3 \pm 0.8$ \\
LD\#4 & $20.1 \pm 1.2$\\
\hline
\end{tabular}
\label{tab:results}
\end{table}

The achieved results clearly confirm the benefic effects of the SiO$_2$ coating and show the efficacy of the procedure here adopted to investigate this element of the light-detector production protocol.   

\subsection{Energy calibration and light yield of the ZnSe sources}
\label{sec:ec}

\begin{figure}[h!]
\centering
\includegraphics[width=0.6\textwidth]{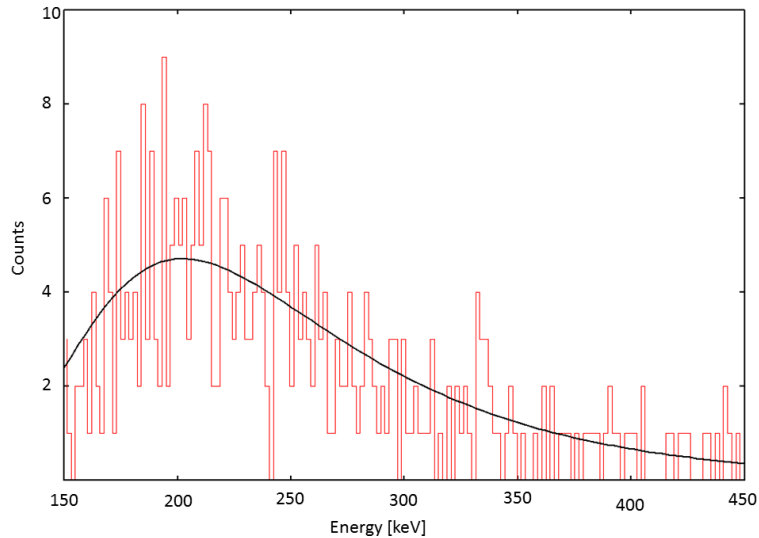}
\caption{(Color online) Amplitude spectrum of the events in detector LD\#3 corresponding to high-energy cosmic muons crossing this detector  and simultaneously an adjacent one. The spectrum is fitted with a Landau distribution in the approximation of Moyal.}
\label{fig:landau}
\end{figure}

The detector energy calibration is not strictly necessary to evaluate the improvement determined by the coating. However, this operation is useful to evaluate the detector energy sensitivity and the source light yield. Unfortunately, only one detector (LD\#3) had a threshold low enough to detect the X-rays emitted by the source described in Section~\ref{sec:sou}. In the spectra acquired with this detector, the energy scale was fixed using the main line given by this source at 5.9~keV. This allowed to determine the position in energy of the bump due to the cosmic muons crossing almost vertically the Ge slab. The same bump was clearly observable in the other three detectors, enabling therefore their energy calibration as well. The position of the bump was obtained by fitting the spectrum in the cosmic muon region with a Landau distribution in the approximation of Moyal~\cite{Moyal}. A good definition of this bump can be obtained by selecting the events in coincidence between two detectors generated by high-energy muons crossing both devices. An example of the spectrum obtained with this coincidence approach for LD\#3 is shown in Fig.~\ref{fig:landau}. The maximum of the fitting curve corresponds to an energy of 180 keV using the X-ray calibration, in good agreement with the expected energy loss of a minimum ionizing particle. 

The energy calibration of the detectors allows to determine the relative scintillation light yield of each ZnSe slab for alpha particle excitation. This can be defined as the ratio between the energy (determined thanks to the absolute calibration of the Ge detector) corresponding to the maximum of the main semi-gaussian used to fit the light spectrum and the main alpha energy, which is about 4.2~MeV. Of course this approach does not take into account geometrical and intrinsic light collection efficiency. The results are collected in Table~\ref{tab:yield}. The relative light yield is reported in case of bare and coated light detector. We have also reconstructed a light yield closer to its expected absolute value assuming that the reflected light is lost to detection, following the arguments exposed in Section~\ref{sec:coat}.

The relative light yield of the crystalline samples from which the source were obtained were measured in another work~\cite{dafinei}, with light detectors similar to ours. The present results are consistent with the previous measurements. The remarkable variability of the scintillation light yield among different ZnSe samples, already observed in other occasions, is confirmed here. 

\begin{table}[h]
\caption{Relative light yield (LY) of the ZnSe slabs used as light sources, measured using the bare and the coated side of the Ge light absorbers. The reconstructed light yield (by taking into account the reflected light fraction) is reported as well. The light yields scale as those of the crystalline samples from which the sources were obtained and measured elsewhere~\cite{dafinei}. The identification number (ID no.) assigned at each source corresponds to that used in reference~\cite{dafinei}.}
\centering
\begin{tabular}{|c|c|c|c|c|}
\hline
Source & Detector & LY (bare Ge) & LY (coated Ge) & Reconstructed LY \\
(ID no.) & & [keV/Mev] & [keV/Mev] & [keV/Mev] \\
\hline
\#8 & LD\#4 & 14.7 & 17.7 & 28.6 \\
\#8 & LD\#1 & 14.4 & 17.1 & 28.0 \\
\#4 & LD\#3 & 12.1 & 14.6 & 23.5 \\
\#1 & LD\#2 & 11.7 & 14.0 & 22.8 \\
\hline
\end{tabular}
\label{tab:yield}
\end{table}

% LD2 -> LT5 - 2nd from above - S 1.3 -> coating in Run1 - the warmest
% LD1 -> LT6 - 1st from above - S 8.1 -> coating in Run1 - the second coldest
% LD3 -> LT3 - 3rd from above - S 4.3 -> coating in Run2 - the best and the coldest
% LD4 -> LT4 - 4th from above - S 8.2 -> coating in Run2 - the second warmest (varying)

\section{Conclusions and prospects}

The set-up described in this paper has allowed to measure unambiguously the positive effect induced by a SiO$_2$ coating on the Ge~absorbers of bolometric light detectors in terms of light collection. The measurement was performed at very low temperatures and in conditions very similar to those in which scintillating bolometers normally operate. The increase of the collected light can be explained by the simple model of refraction index matching. 

Since the refraction index of SiO$_2$ is far from the value that minimizes the fraction of the reflected light (which is around 2.4), we plan to test the effect of other coating materials in the same set-up. Interesting candidates are SiO (n=2.5), ZnS (n=2.6) and TiO$_2$ (n=2.5)~\cite{ri}, which should increase the absorbed light fraction by $\sim 35$\% with respect to bare germanium, to be compared with the $\sim 20$\% value measured in this work. This improvement (together with other measures aiming at increasing the energy sensitivity of the light detectors) would be very useful for the exploitation of the Cherenkov effect in TeO$_2$ as a method to reject the alpha background~\cite{Taba}. Of course, other more sophisticated forms of antireflective measures such as interference and textured coatings may be taken into consideration.

The work here described was performed within the project LUCIFER, funded by the European Research Council under the EU Seventh Framework Programme (ERC grant agreement n. 247115). We thank Ioan Dafinei for providing us with the ZnSe slabs used to make the light sources.

\end{document}